\documentclass[fleqn,usenatbib]{mnras}

\usepackage{newtxtext,newtxmath}

\usepackage[T1]{fontenc}

\DeclareRobustCommand{\VAN}[3]{#2}
\let\VANthebibliography\thebibliography
\def\thebibliography{\DeclareRobustCommand{\VAN}[3]{##3}\VANthebibliography}

\usepackage{graphicx}	

\usepackage{soul}
\usepackage[dvipsnames]{xcolor}

\title[BLAPs in BPASS]{Binary evolution pathways of Blue Large-Amplitude Pulsators}

\author[Byrne, Stanway \& Eldridge]{
C. M. Byrne,$^{1}$\thanks{E-mail: conor.byrne@warwick.ac.uk (CMB)}
E. R. Stanway,$^{1}$ and
J. J. Eldridge$^{2}$
\\
$^{1}$Department of Physics, University of Warwick, Gibbet Hill Road, Coventry, CV4 7AL, UK\\
$^{2}$Department of Physics, University of Auckland, Private Bag 92019, Auckland, New Zealand
}

\date{Accepted XXX. Received YYY; in original form ZZZ}

\pubyear{2021}

\begin{document}
\label{firstpage}
\pagerange{\pageref{firstpage}--\pageref{lastpage}}
\maketitle

\begin{abstract}
Blue Large-Amplitude Pulsators (BLAPs) are a recently discovered class of pulsating star, believed to be proto-white dwarfs, produced by mass stripping of a red giant when it has a small helium core. An outstanding question is why the stars in this class of pulsator seem to form two distinct groups by surface gravity, despite predictions that stars in the gap between them should also pulsate. We use a binary population synthesis model to identify potential evolutionary pathways that a star can take to become a BLAP. We find that BLAPs can be produced either through common envelope evolution or Roche lobe overflow, with a Main Sequence star or an evolved compact object being responsible for the envelope stripping. The mass distribution of the inferred population indicates that fewer stars would be expected in the range of masses intermediate to the two known groups of pulsators, suggesting that the lack of observational discoveries in this region may be a result of the underlying population of pre-white dwarf stars. We also consider metallicity variation and find evidence that BLAPs at $Z=0.010$ (half-Solar) would be pulsationally unstable and may also be more common. Based on this analysis, we expect the Milky Way to host around 12000 BLAPs and we predict the number density of sources expected in future observations such as the Legacy Survey of Space and Time at the Vera Rubin Observatory.
\end{abstract}




\section{Introduction}
\label{sec:intro}

Blue Large-Amplitude Pulsators (BLAPs) are hot, faint stars, discovered in both OGLE and ZTF survey data \citep{Pietrukowicz17,Kupfer19}. They show periodic brightness variations on timescales ranging from 3 minutes to 40 minutes, with amplitudes of up to a few per cent of their total brightness, and effective temperatures $(T_{\rm{eff}})$ between $20\,000$ and $35\,000\,\rm{K}$. Stellar evolution modelling suggests they are likely low-mass pre-white dwarf stars (pre-WDs, $\sim 0.2-0.4\,\rm{M}_\odot$), formed following significant mass loss from a red giant star in a binary system, either through common envelope evolution or Roche Lobe overflow \citep{Romero18,Byrne18b}. 

The pulsation driving mechanism in BLAPs is the $\kappa$-mechanism, due to the opacity bump generated by iron and nickel in the envelope. An enhancement in the abundance of these elements - produced by the equilibrium between radiative levitation and gravitational settling processes - is necessary to produce a sufficient opacity bump to drive the pulsations \citep{Charpinet97,Byrne20}. It has also been suggested that BLAPs could be formed from the surviving remnant of a single degenerate Type Ia supernova \citep{Meng20} or that they could be core-helium burning stars approaching the extended horizontal branch \citep{Wu18}. 

Low-mass pre-white dwarfs are lower mass counterparts to core-helium burning hot subdwarfs \citep[For a detailed review of hot subdwarfs, see][]{Heber16}. Some hot subdwarfs are also known to pulsate, however the amplitude of their pulsations is typically much smaller than that seen in BLAPs.

Eighteen BLAPs have been reported in the literature, 14 through the Optical Gravitational Lensing Experiment survey of the Galactic Bulge \citep[OGLE,][]{OGLE} with periods of 20-40 min \citep{Pietrukowicz17} and 4 from a high-cadence survey of the Galactic Plane by the Zwicky Transient Factory \citep[ZTF,][]{ZTF} with periods between 3 and 8 min \citep{Kupfer19}. Evolutionary tracks of low-mass pre-white dwarf star models indicate that these two groups of objects have masses of $\,\sim0.30\,\rm{M}_\odot$ and $\,\sim0.26\,\rm{M}_\odot$ respectively \citep{Byrne20}).

Even though \cite{2018A&A...620L...9R} showed that BLAPs can be identified using Gaia photometry and parallax, spectroscopic observations are necessary to assess further properties of the BLAP population.
 
Currently 8 of the 18 BLAPs have spectroscopic observations, 4 from the ZTF \citep{Kupfer19} and 4 from OGLE \citep{Pietrukowicz17}). In the case of the BLAPs from the OGLE survey, the determinations of surface gravity are between $\log(g/\rm{cm}\,\rm{s}^{-2})=4.2$ and 4.6, while for the ZTF pulsators, $\log(g)$ ranges from 5.3 to 5.7. These are referred as `low-gravity BLAPs' and `high-gravity BLAPs', respectively \citep{Kupfer19, Byrne20}.

\cite{Byrne20} computed a set of detailed stellar evolution models including atomic diffusion with radiative levitation for a sequence of low-mass pre-WD stars, produced via hydrogen envelope stripping on the Red Giant Branch. Non-adiabatic pulsation analysis found that low-mass, pre-WD stars between $\sim 0.25\,\rm{M}_\odot$ up to $0.45\,\rm{M}_\odot$ are unstable in the radial fundamental mode while in this region of $\log(g)-\log(T_{\rm{eff}})$ space. 

This contiguous region of instability suggests that the gap in $\log(g)$ does not result from a lack of pulsation driving, leaving two plausible explanations; a physical gap owing to a dearth of low-mass pre-WDs in the intermediate mass range, or observational bias in detecting faint blue stars with periods between $8$ and $20\,$min. 

In this study, we use a binary population synthesis model to identify formation channels which produce low-mass pre-WDs, and are thus likely to behave like BLAPs. 
Our objectives are four-fold: (i) We  consider whether additional evolution pathways exist that might give rise to stars matching the known properties of BLAPs; (ii) we determine the likely number of BLAPs in the Milky Way - a quantity which has ranged over three orders of magnitude in previous estimates \citep{Meng20}; (iii) we investigate whether the observed mass gap in known systems arises naturally in a synthetic population; (iv) we calculate the expected surface density of these sources in deep, wide-field surveys. Our motivation is to provide a framework for contextualising these sources within the wider stellar population. As the observed sample of BLAPs grows, detailed comparison with such models may yield independent insights into the binary fraction, metallicity and star formation history of the Galactic disk population and into the formation pathway of low mass white dwarfs.

The structure of this paper is as follows: The binary population synthesis methodology is introduced in section \ref{sec:methods}. We consider the likelihoods of each pathway and the amount of time individual stellar models spend in this evolutionary phase in section \ref{sec:cands}. This allows us to explore the properties of the population as a function of mass and time. In section \ref{sec:stats} we build on this analysis to estimate the number of these stars that are present in the Galaxy. In section \ref{sec:discussion} we discuss the implications and interpretation of our results, and consider observational constraints to predict the expected number of BLAPs which may be observable by the Legacy Survey of Space and Time (LSST) at the Vera Rubin Observatory. In section \ref{sec:uncertainties} we discuss uncertainties on the above predictions, before presenting our conclusions in section \ref{sec:conc}.

\section{Population Modelling}
\label{sec:methods}

\subsection{BPASS}
\label{sec:bpass}

Binary Populations and Spectral Synthesis \citep[BPASS, ][]{2017PASA...34...58E} is a detailed stellar population synthesis code, which utilises a custom grid of stellar evolution models, originally based on the Cambridge STARS code \citep{Eggleton71,Pols95,Eldridge04,Eldridge08}. Stars are evolved in one dimension using a shellular approximation. In addition to single stars, BPASS also models binary evolution. Mass and angular momentum can be transferred between primary and secondary in a binary (the initially more massive and less massive components of the binary respectively), or lost from the system entirely. Evolution of the binary parameters, as well as the stellar structure, is tracked. BPASS primary models are constructed on a grid of initial primary star masses between 0.1 and 300\,M$_\odot$, periods between 1 and $10^4$ days and mass ratios between 0 and 0.9, as well as 13 metallicities ranging from $Z=10^{-5}$ to 0.04. Binaries with an initial period greater than $10^4$ days are unlikely to interact significantly and are treated as single. A grid of secondary models supplement the primary models and continue to evolve the initially lower mass stars after the collapse of the primary into a compact object.

Population synthesis is carried out through the weighted combination of primary models. The population weightings are normalised to give a total stellar mass of $10^6$\,M$_\odot$. Weights are applied based on a joint probability distribution:

\begin{equation*}
  W_\mathrm{pop}(M_1,P,q) = \xi(M_1) \, f_\mathrm{bin}(M_1) \, f_\mathrm{log P}(M_1) \, f_q(M_1, \log P),
\end{equation*}

where $\xi(M_1)$ is initial mass function (IMF) from which stars are drawn as a function of the mass of the primary $M_1$, $f_\mathrm{bin}$ is initial binary fraction as a function of $M_1$, $f_\mathrm{log P}$ is the initial binary period ($P$) distribution also as a function of $M_1$, and $f_q$ is initial mass ratio ($q)$ distribution as a function of both the primary mass and the initial period of the binary. In BPASS v2.2.1 \citep{2018MNRAS.479...75S}, the adopted initial binary parameter distributions are based on the empirical analysis of \citet[see their table 13]{2017ApJS..230...15M}. We interpolate linearly between the multiple star fractions estimated by those authors in five primary mass bins, and three bins of initial period at each mass. The initial mass ratio distribution is described by a broken power law slope $p_q \propto q^\alpha$ where $\alpha$ is again constrained empirically in each mass and period bin, and the distribution is interpolated between bins. In addition, the expected excess fraction of near-twin systems (i.e. $q\sim1$) is assigned to the highest mass ratio models available in BPASS ($q=0.9$).  The full implementation of population synthesis in BPASS is explained in detail in \citet{2017PASA...34...58E} and \citet{2018MNRAS.479...75S}, and the uncertainties arising from the adopted empirical companion frequencies are explored in \citet{2020MNRAS.495.4605S,2020MNRAS.497.2201S}.

At the end of the life of the primary, a probabilistic step determines the likelihood of a binary being disrupted, and the expected masses and period of surviving binaries, given a distribution of asymmetric supernova kicks. These probabilities provide a weighting for models on a secondary grid, with several primary models potentially contributing to the likelihood weighting of a single secondary model. In a small number of cases, the result of the primary evolution is a stellar merger, or rapid mass transfer which results in rapid mixing and rejuvenation of the secondary star. These cases are also tracked in the models, as is the eventual merger of double-compact binaries through gravitational wave radiation driven inspirals.

The adopted IMF, $\xi(M_1)$, is a broken power law, with a slope of -0.30 between 0.1 and 0.5 M$_\odot$ and a slope of -1.35 between 0.5 and 300 $M_\odot$. This initial mass function is similar to that proposed by \citet{2001MNRAS.322..231K}. The flattened slope at low masses accounts for the deficit of low mass stars observed in the Milky Way disk compared to that expected from an unbroken power law and produces a similar mass distribution to the \citet{2003PASP..115..763C} IMF at $M<1$\,M$_\odot$ \citep[as demonstrated in][]{2018MNRAS.479...75S}. The upper mass slope \citep[which matches that of ][]{1955ApJ...121..161S} is consistent with a range of Galactic and extra-galactic observations  \citep[see][for a discussion of the uncertainties associated with IMF estimates]{2018PASA...35...39H}. The upper mass limit is motivated by the observation of very massive stars ($M\sim100-300$\,M$_\odot$) in young, massive star formation regions such as those identified in the Tarantula Nebula \citep[e.g.][]{2020MNRAS.499.1918B}. Adopting a different upper mass limit has a negligible effect on the low mass stellar population considered in this work.

BPASS is extensively used for interpretation of massive star populations \citep[e.g.][]{2021MNRAS.tmp.1487P,2021MNRAS.tmp.1313C,2021MNRAS.501.3289V,2021ApJ...908...87N,2020MNRAS.499.3819M,2020MNRAS.498.1347S,2020ApJ...900..118N,2020ApJ...899..117S,2020MNRAS.495.4430T,2020MNRAS.495.1501C,2020ApJ...896..164D,2020MNRAS.493.6079W,2020MNRAS.491.3479C,2020ApJ...888L..11S}. Its accuracy and capabilities in the low mass regime have  been explored through the white dwarf progenitor initial to final mass ratio \citep{2017PASA...34...58E}, the impact of low mass binary stars on understanding ages of globular clusters and elliptical galaxies \citep{2018MNRAS.479...75S} and consideration of the type Ia supernova rate \citep{2020MNRAS.493L...6T}, producing good matches to the data in each case. As a relatively short-lived evolution phase strongly dependent on mass transfer, the BLAP population provide an interesting low mass test case, and a science application for which BPASS population synthesis can provide new insights.

\subsection{Selecting models of interest}
\label{sec:selecting}

Stellar structure and evolution models at Solar metallicity ($Z=0.020$) have been searched for candidate models with properties which resemble those of BLAPs at any time, $t$, during their evolution\footnote{Analysis of these models made use of the {\texttt{pandas}} data frames of the BPASS models \citep{BPASS_pickles}, which were constructed by the {\sc{hoki}} Python package \citep{HOKI}}. 

We select models which satisfy:
\begin{enumerate}
    \item $4.4 \le \log\left({T}_{\rm{eff}}/\rm{K}\right) \le 5.5$ 
    \item $3.5 \le \log(g/\rm{cm}\,\rm{s}^{-2}) \le 6.0$ 
    \item $t \le 1.4\times10^{10}\,\rm{yr}$ 
    \item $ M \le 0.5\,\rm{M}_{\odot}$ at time $t$.
    \item He-rich core at time $t$
\end{enumerate}

\begin{figure}
    \centering
    \includegraphics[width=0.48\textwidth]{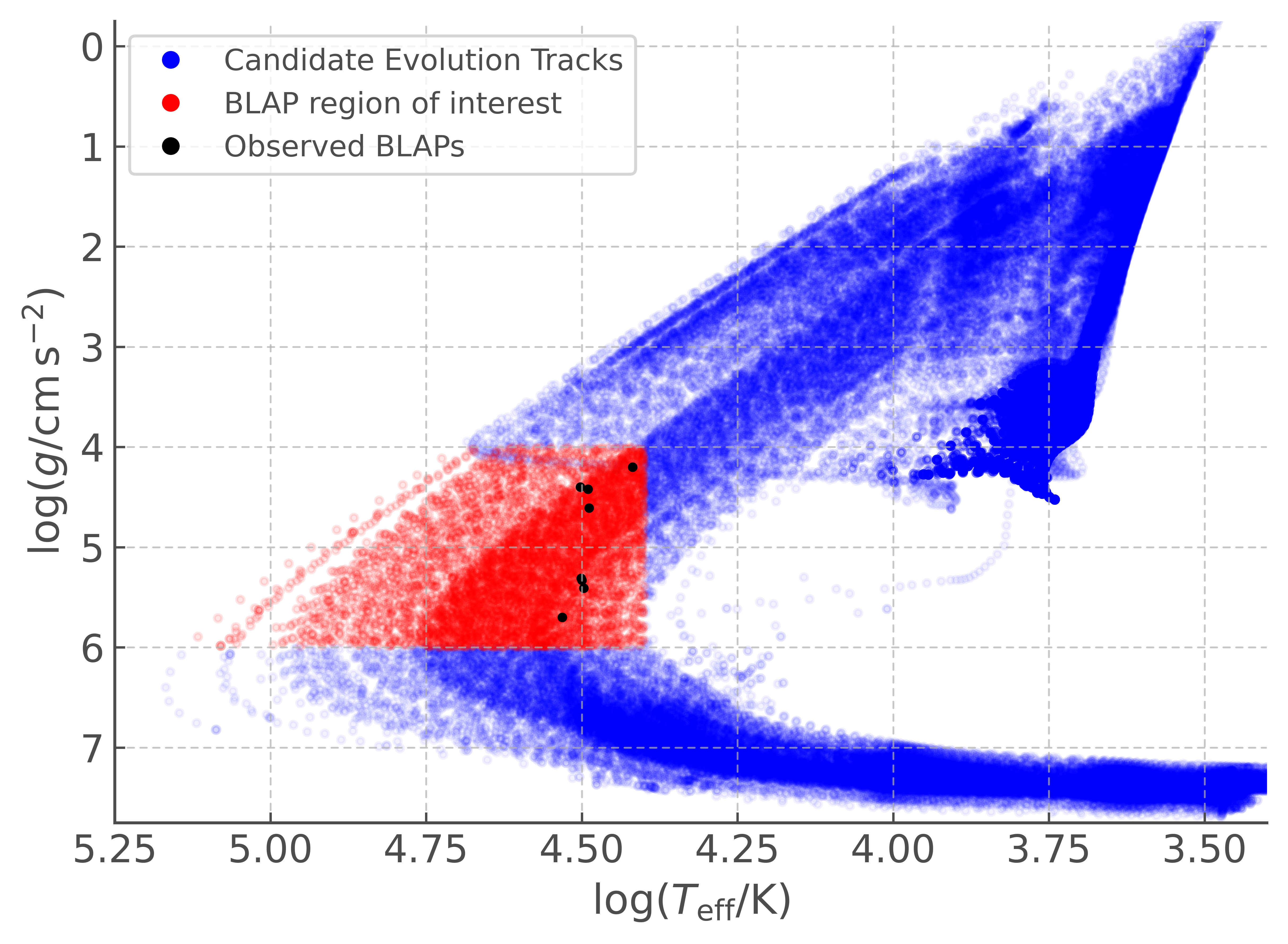}
    \caption{Evolutionary tracks of BPASS models which satisfy our criteria as potential BLAP models.}
    \label{fig:filtered_models}
\end{figure}

Constraints (i) and (ii) arise from the spectroscopic properties of observed BLAPs and the location of the instability region identified by \cite{Byrne20}. Constraint (iii) requires stars to have formed within the current lifetime of the Universe. Constraint (iv) is predicated on the assumption that we are considering low-mass pre-white dwarfs; stars more massive than this will not match that definition. Constraint (v) excludes models which become hot subdwarfs, or otherwise form a CO-rich core. As \cite{Kupfer19} have shown, the expected period of the radial fundamental mode in a pre-He WD is shorter than a hot subdwarf/helium-burning star at the same point in the spectroscopic HR diagram, providing a means of distinguishing these two structures, and a contracting pre-He WD provides a much better match to the observed pulsation period of their objects than a more massive CO core object. Hence, we remove the CO core models from our sample, and focus solely on the objects which are pre-He WDs hereafter. While CO core stars that are towards the end of their core-helium-burning lifetime may also display BLAP-like behaviour, the fate of such objects needs further study.

This selection identifies models which pass through the instability region of \cite{Byrne20}, with no direct knowledge of their pulsation properties. Radiative levitation is expected to produce an opacity bump capable of driving pulsation if the star spends an extended time in this region. Nonetheless, it is important to emphasise that a complete pulsation analysis of these candidate models (hereafter referred to as BLAP models) has not been carried out. Pulsation analysis of these BLAP models at non-solar metallicities have also not been examined in detail in this work, however an initial exploratory model suggests that stars at a half-Solar metallicity do indeed become pulsationally unstable. Further work is clearly needed in this area.

Fig~\ref{fig:filtered_models} illustrates the evolution tracks of stellar models in BPASS which satisfy all constraints on age, mass, effective temperature and surface gravity. The region highlighted in red is the region of parameter space where constraints (i) to (iv) are satisfied. The black points indicate the properties of observed BLAPs which have spectroscopic measurements. A small number of high mass ($3.2 - 50 \,\rm{M}_\odot$) stars can produce BLAP models by these criteria. These models are not plotted on the figures, but they will be briefly discussed in Section~\ref{sec:himass}. The selected models all proceed from core hydrogen burning on the Main Sequence, to envelope stripping on the red giant branch, before cooling to become a white dwarf. 

A total of 754 BPASS stellar evolution models satisfy the selection criteria at some period in their evolution. 

\section{Properties of the BLAP models}
\label{sec:cands}

\subsection{Binary pathway}
\label{sec:pathway}

Of the models selected, 485 are primary stars, while 269 are initially secondary stars; none are from single evolution pathways. However 46 models become single prior to reaching the BLAP phase, their companion having undergone a type Ia supernova  \citep[as also identified by][]{Meng20}. This is unsurprising, as it takes a time significantly longer than the current age of the Universe to form hot subdwarfs and low mass white dwarfs through single star evolution \citep[e.g.][]{ClausenWade}.

\begin{figure}
    \centering
    \includegraphics[width=0.48\textwidth]{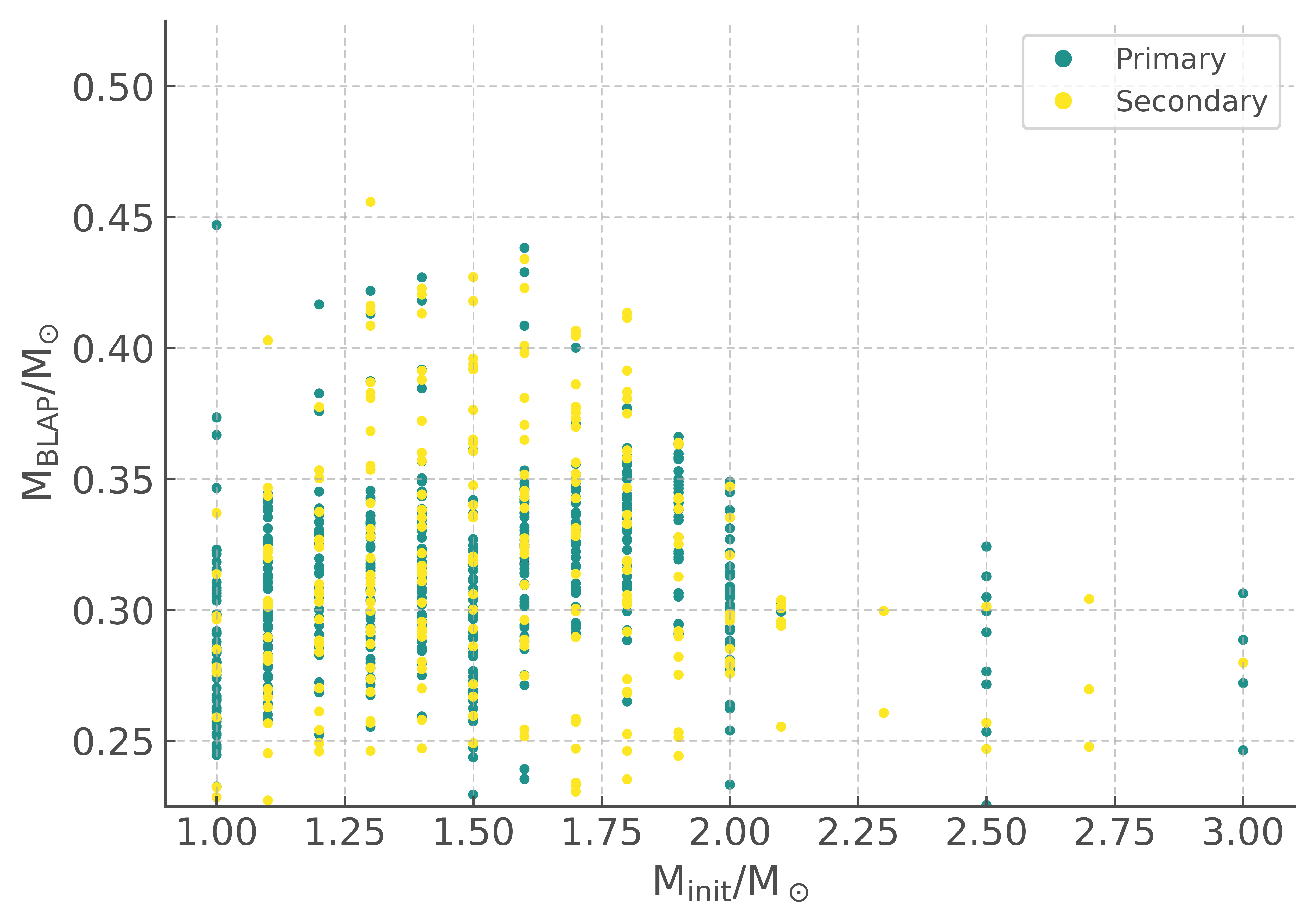}
    \caption{Distribution of initial mass and BLAP mass of the selected BPASS models, coloured according to the type of binary system which produced them. The x-axis  in this plot (and subsequent plots) has been truncated for clarity, but 13 further models exist with initial masses between 3.2 and 60 $\mathrm{M}_\odot$}
    \label{fig:model_type}
\end{figure}

\subsection{Initial Masses}
\label{sec:init_m}

Fig~\ref{fig:model_type} compares the distribution of initial masses with the resulting mass of the BLAP models, coloured by model type. The BLAP mass is defined as the arithmetic mean of the stellar mass of a model when it enters and leaves the selected region. Green points indicate models which are the primary component of their system, while yellow points indicate stars which were initially the secondary component in their binary. There is no strong correlation between initial and final mass, indicating that binary parameters such as mass ratio and initial separation are more important factors in determining the final mass of the BLAP models. 

A broad range of initial masses can produce BLAP models, given their narrow range of final masses. Although most progenitors (709 individual models) have an initial mass between 1 and 2$\,\rm{M}_\odot$, a number of more massive progenitors (45 models) also exist. The minimum progenitor initial mass is $1.0\,\rm{M}_\odot$, a constraint provided by the age of the Universe. The minimum BLAP mass is $\sim0.23\,\rm{M}_\odot$, as less massive pre-WDs will not reach sufficiently high maximum temperatures to enter the region of interest.

The distribution of objects in initial mass versus BLAP mass (Fig~\ref{fig:model_type}) excludes objects with initial masses between 1 and 2$\,\rm{M}_\odot$ and final masses of around $0.47\,\rm{M}_\odot$, which would be a typical expected pathway for a hot subdwarf star. 

\begin{figure}
    \centering
    \includegraphics[width=0.48\textwidth]{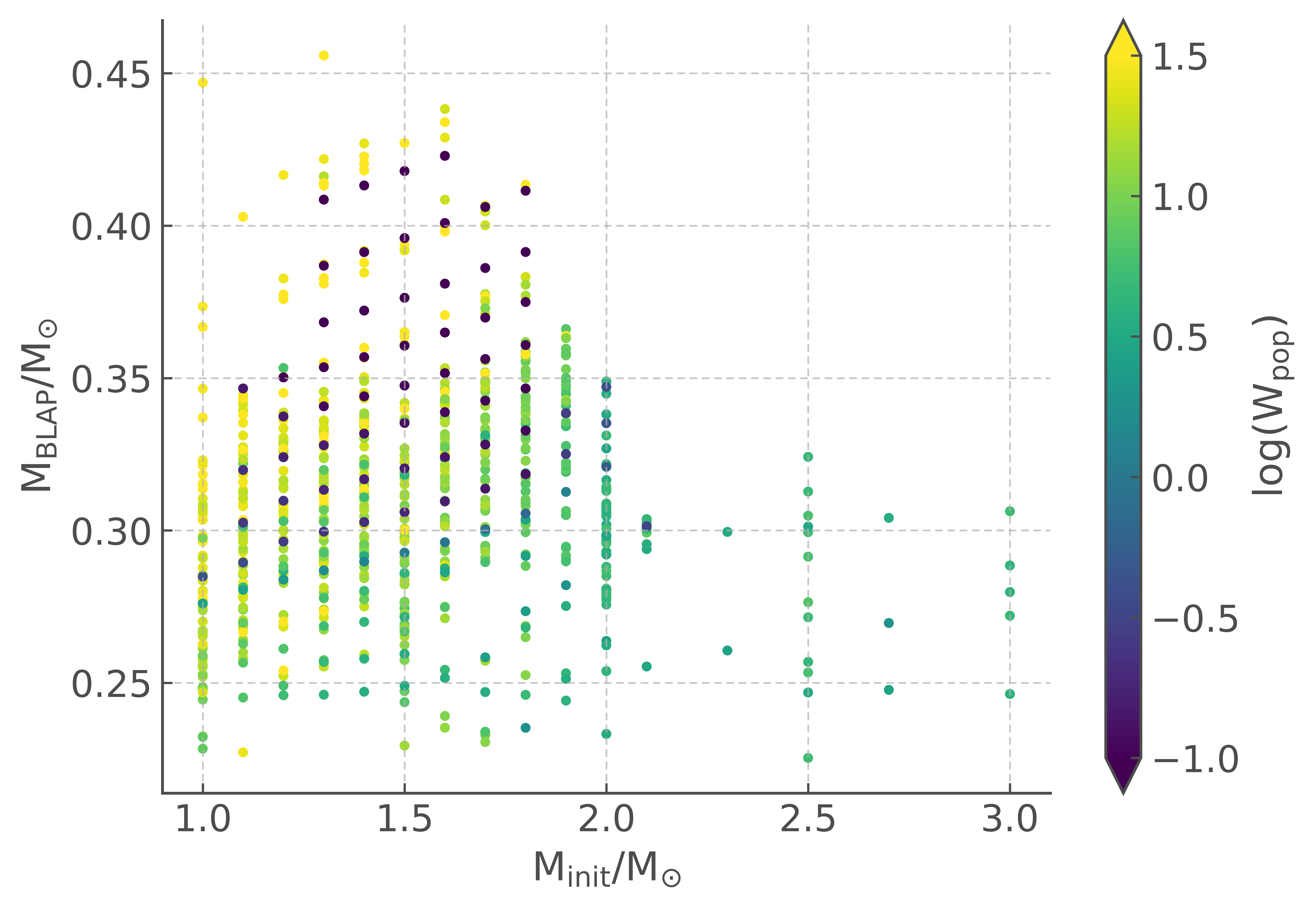}
    \caption{Initial mass--BLAP mass relationship, with colour coding to indicate the population synthesis weighting of each individual model.}
    \label{fig:m_init_m_blap_imf}
\end{figure}

\subsection{Pathway weightings}
\label{sec:pathway_wt}

Determining the relative likelihoods of any individual BLAP model formation channel requires the use of the weightings, $W_{\rm{pop}}$, attached to each individual model in the population synthesis, as outlined in Section~\ref{sec:bpass}. Fig~\ref{fig:m_init_m_blap_imf} illustrates the population synthesis weighting, $W_{\mathrm{pop}}$ of each model selected. 

Figure~\ref{fig:m_init_m_blap_imf} shows a number of models at fixed initial mass which vary in population synthesis weighting, particularly noticeable for high BLAP-mass models with M$_\mathrm{init}<2$\,M$_\odot$. This is a consequence of considering a range of binary initial mass ratios. The lower weighted models have massive, compact objects as companions. This means the BLAP progenitor initially had a much more massive companion, and is disfavoured by the initial mass function.

There is also a weak trend in population synthesis weighting towards more frequent massive BLAP models at fixed initial mass. This arises since systems with wide initial separations are more common than close binaries. More massive BLAP models will form from systems with wider initial orbits, as these stars will evolve further up the red giant branch, amassing a larger helium core before undergoing mass transfer.

\begin{figure*}
    \centering
    \includegraphics[width=0.85\textwidth]{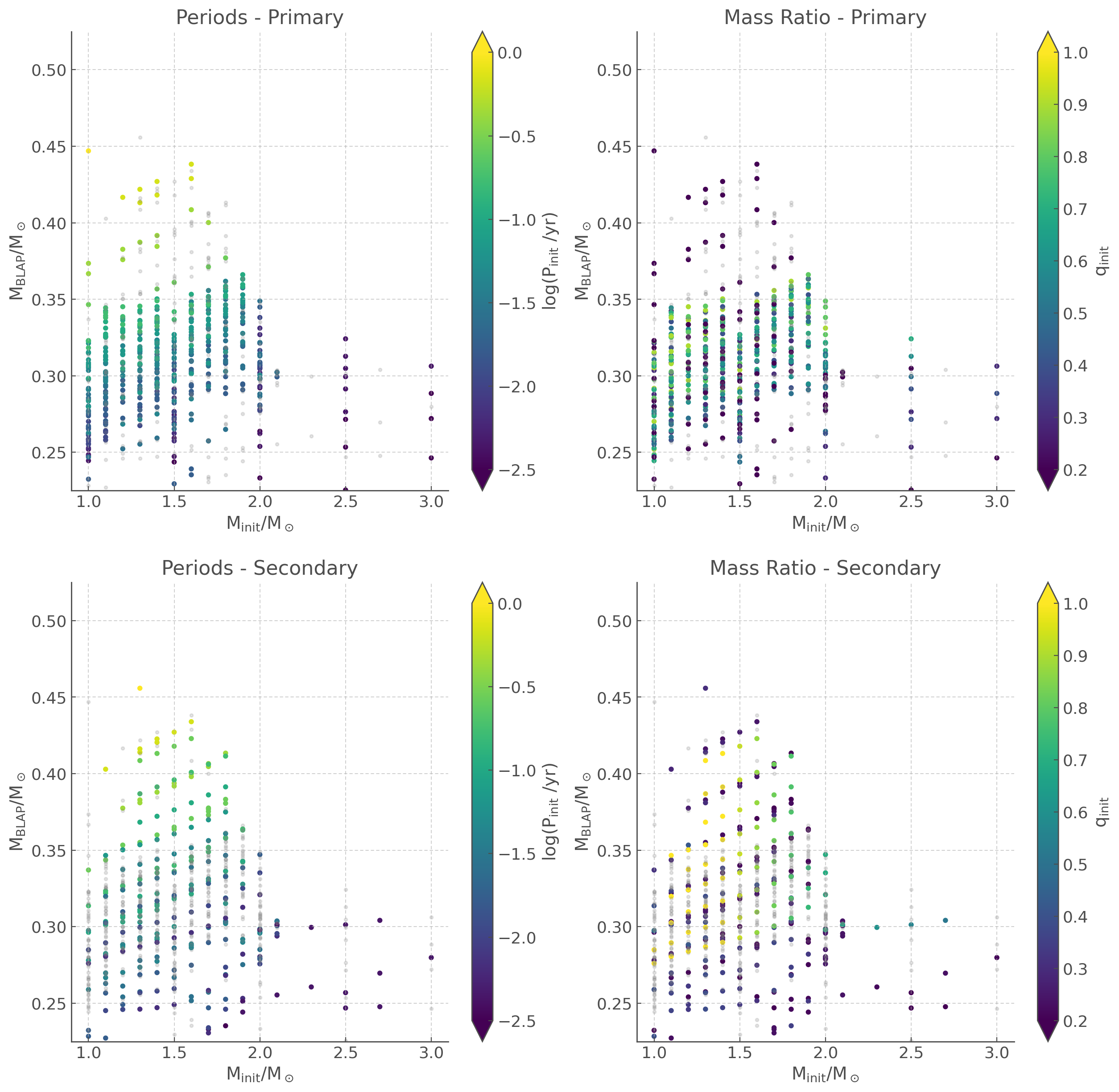}
    \caption{Initial mass -- BLAP mass relationship colour-coded by initial orbital period (left panels) and mass ratio (right panels), separated into primary (top panels) and secondary (bottom panels). In each plot, the grey models show the locations of the secondary/primary models.}
    \label{fig:pinit_qinit}
\end{figure*}

Fig~\ref{fig:pinit_qinit} illustrates the orbital periods and mass ratios of systems that produce the models of interest, again distributed in initial mass -- BLAP mass parameter space. Outliers above $3\,\rm{M}_\odot$ (4 primary models and 9 secondary models, after removal of systems with CO cores) are not shown in these plots as they make very little impact on the trends and patterns in the data. The initial orbital periods for BLAP models (upper left panel) show a trend of increasing orbital period with increasing resultant BLAP mass. This is consistent with the evolutionary picture of these stars: a star in a system with a larger initial orbital period will evolve for longer on the RGB before expanding sufficiently to transfer mass via Roche Lobe overflow or common envelope evolution. This will result in the star having a more massive helium core, thus producing a more massive pre-WD after the mass transfer phase. 

This behaviour is less clear in the secondary models (lower left panel), where there is more scatter in initial orbital periods. The initial period and mass is defined at the end of the life of the primary, hence this is probably a result of prior evolution in the system. If there was an initial mass transfer phase during the lifetime of the primary, then the `initial' orbital period as indicated by this model will be shorter, and the evolution of the secondary (the BLAP progenitor) may have been rejuvenated by such a mass transfer event. 

There are no clear trends in the initial binary mass ratios (right hand column), although there is a weak indication of decreasing mass ratio with increasing initial mass in 
a subset of the secondary models. 

\subsection{Lifetimes}
\label{sec:lifetimes}

The lifetimes of these stars in the BLAP region of interest (left column) vary from $10^3\,$yr up to almost $10^7\,$yr. More massive pre-WD stars cross the region in a shorter time period. This is consistent with the findings of \cite{Byrne20} and a consequence of the more massive post-RGB stars contracting faster due to their short nuclear timescales for residual shell-hydrogen burning. 

The age of the BLAP when it reaches this region of interest (right column) varies from around $10^{8.5}\,$yr to $10^{10.1}\,$yr, and is strongly correlated with the initial mass. This is unsurprising as the evolution timescale is dominated by the Main Sequence lifetime, which is inversely proportional to mass.

Some models show much shorter `BLAP lifetimes' than the general trends. These models correspond to those left with the smallest hydrogen envelopes after mass transfer. They evolve through the region much more rapidly as they have very little material to sustain hydrogen shell burning to slow their contraction.

\begin{figure*}
    \centering
    \includegraphics[width=0.85\textwidth]{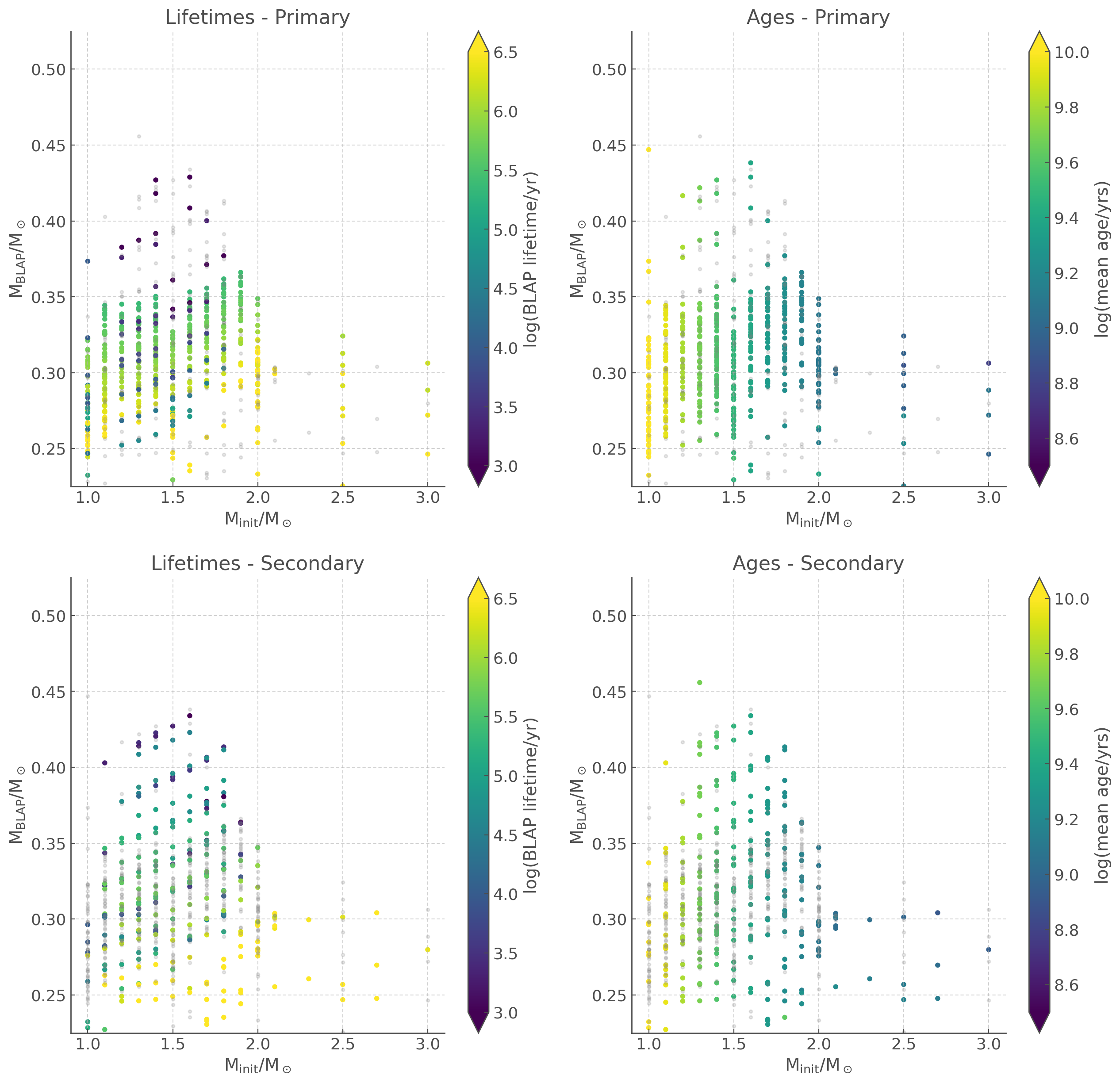}
    \caption{As in Fig~\ref{fig:pinit_qinit}, but colour-coded according to lifetime in the BLAP region of interest (left panels) and the age of the model when it is in the region (right).}
    \label{fig:age_life}
\end{figure*}

\section{Statistics}
\label{sec:stats}

\subsection{Number counts}
\label{sec:number_count}
The total number of BLAPs in the Milky Way is important for predicting source counts in future observations and could theoretically help to constrain evolution pathways, if the observational completeness is well understood.
Hence we calculate $N_{\rm{stars}}$, the number of possible BLAPs that each specific model contributes to the Milky Way population. The total amount of material forming stars, $M_{\rm{SF}}$, in a given time interval, $\Delta t$, is given by the assumed star formation rate of the Milky Way.
\begin{equation}
M_{\rm{SF}} = \rm{SFR}\times\Delta t
\end{equation}
\noindent where $\Delta t$ is the width of the time bin being considered. We consider time bins between $10^6\,$yr and $10^{10.2}\,$yr of equal width in log-space. In this work we primarily concern ourselves with a simplified prescription for the Galactic disk population in which the star formation rate (SFR) remains at a constant value of $3\,\rm{M}_{\odot}\,\rm{yr}^{-1}$.

For each of the selected models that enters the BLAP phase during the chosen time bin, multiplying $M_{\rm{SF}}$ by the population synthesis weighting of that model will indicate the proportion of the star forming material that will form stars corresponding to that particular combination of mass and binary parameters, $M_{\rm{SF, BLAP}}$. Dividing $M_{\rm{SF, BLAP}}$ by the mass of the binary system in question gives the number of star systems which will form from this binary evolution pathway. The mass of the binary is determined by multiplying the initial mass of the star by a factor of $1+q$, where $q$ is the initial mass ratio,  $M_{\rm{sec}}/M_{\rm{pri}}$:

\begin{equation}
N'_{\rm{stars}} = \frac{M_{\rm{SF}}\times W_{\rm{pop}}}{(1+q)\rm{M}_{\rm{init}}}.
\end{equation}

\noindent Finally, we account for the fraction of the time bin that the model spends in this phase of evolution,
\begin{equation}
N_{\rm{stars}} = N'_{\rm{stars}}\times \max\left(\frac{t_{\rm{life}}}{\Delta t}\,,\,1\right).
\end{equation}

\begin{figure}
    \centering
    \includegraphics[width=0.48\textwidth]{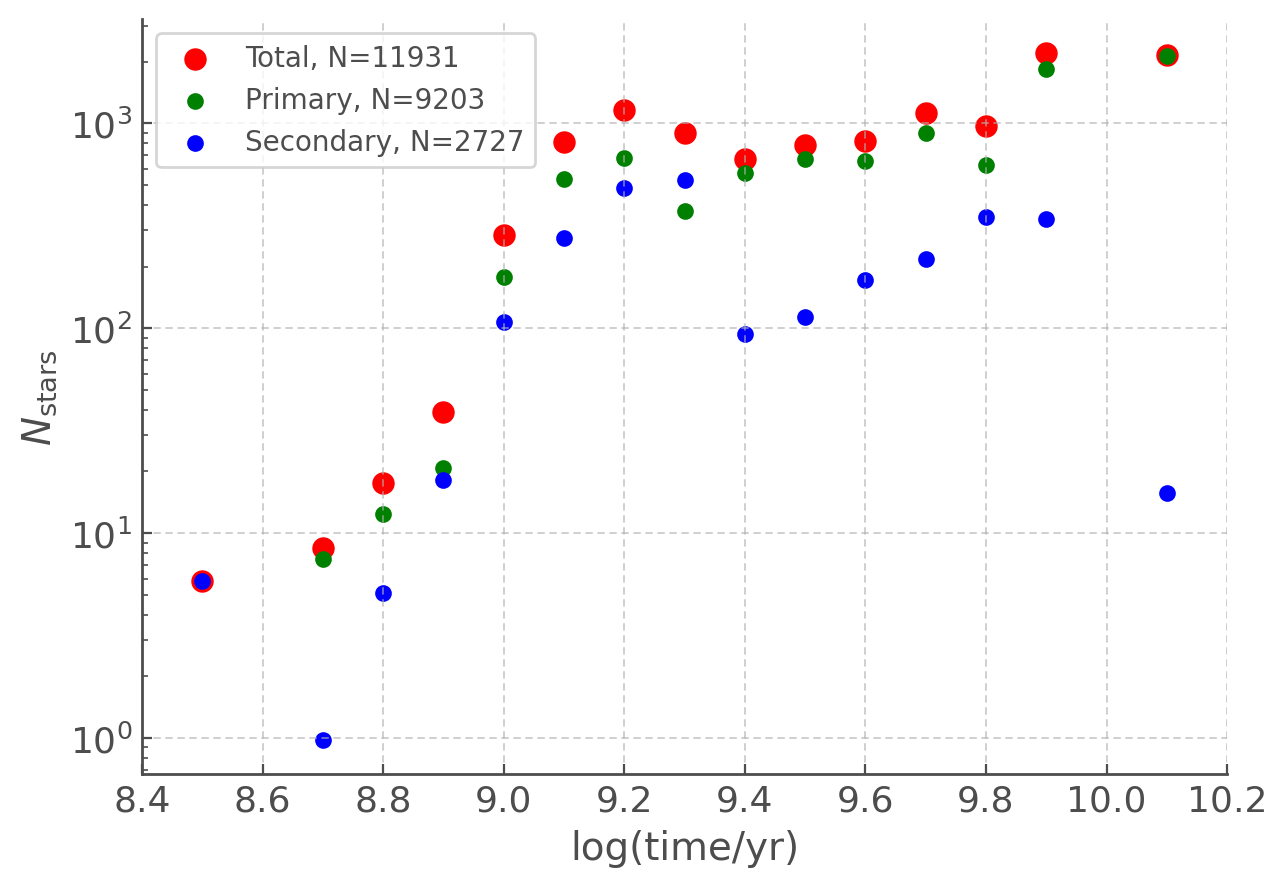}
    \caption{Distribution of number of BLAPs expected in the Milky Way as a function of logarithmic age from initial binary star formation, assuming a constant star formation rate of 3\,M$_\odot$\,yr$^{-1}$.}
    \label{fig:n_stats}
\end{figure}

Here $t_\mathrm{life}$ is the elapsed timespan the BLAP system spends in the region of interest, during this time bin. This provides a total for the number of observable BLAP systems that each individual model is responsible for in a given time bin, for the assumed star formation history. Summing $N_{\rm{stars}}$ over each of the models in each time bin, yields the number of BLAPs forming as a function of age. This is shown in Fig~\ref{fig:n_stats}, where the models have been separated according to whether they are initially primary or secondary stars in their systems. The earliest BLAP models form at around $10^{8.4}\,$yr ($250\,$Myr), with additional systems added to the population at all ages up to $10^{10.1}\,$yr ($12.5\,$Gyr).

The number of primary models is significantly higher than the secondary models. Given our assumed star formation rate, 9\,203 primary stars and  2\,727 secondary stars are expected to be in the BLAP phase at the current time, giving a total of 11\,931 BLAPs scattered across the Milky Way. 

\subsection{Population properties}
\label{sec:pop_prop}

\begin{figure}
    \centering
    \includegraphics[width=0.48\textwidth]{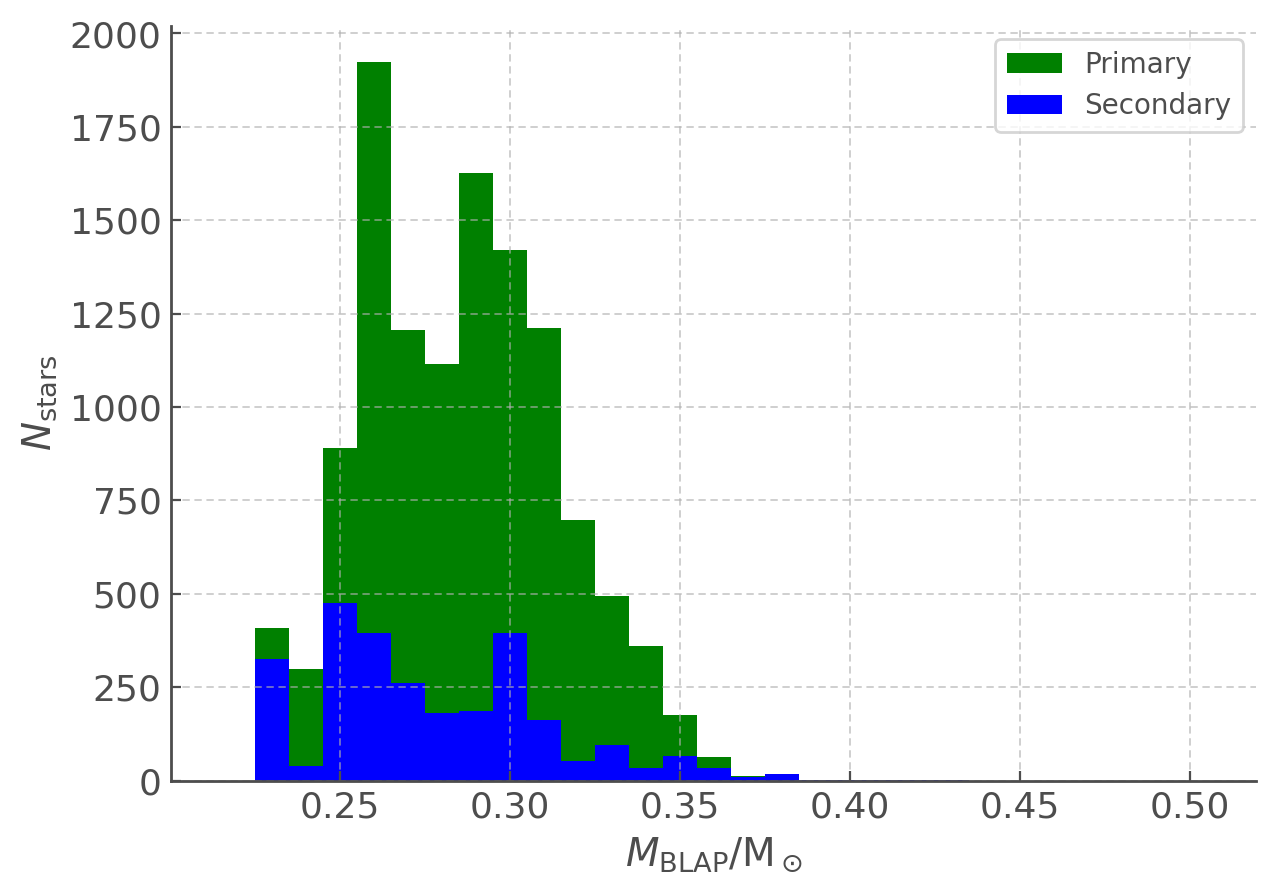}
    \caption{Distribution of number of BLAPs expected in the Milky Way as a function of their mass, in bins of $0.01\,\rm{M}_\odot$, colour coded by stellar model type.}
    \label{fig:mass_distrib}
\end{figure}

Fig~\ref{fig:mass_distrib} demonstrates that the mass distribution of BLAPs has 2 peaks, a sharp peak centred at $0.26\,\rm{M}_\odot$ and a second, broader peak at $0.29\,\rm{M}_\odot$. The observed stars cluster at around $0.26\,\rm{M}_\odot$ and $0.3\,\rm{M}_\odot$, based on matching of pre-WD evolution tracks from \cite{Byrne20} to observed temperatures and gravities reported by \cite{Pietrukowicz17} and \cite{Kupfer19}. This suggests that the observational `gap' between the high-gravity pulsators and their low-gravity counterparts may indeed be a consequence of the underlying mass distribution of the population of low mass pre-WDs, rather than purely an observational selection effect. While the number statistics are low, the figure also suggests that BLAPs arising from a secondary pathway may be biased towards the lower mass category.

\begin{figure}
    \centering
    \includegraphics[width=0.48\textwidth]{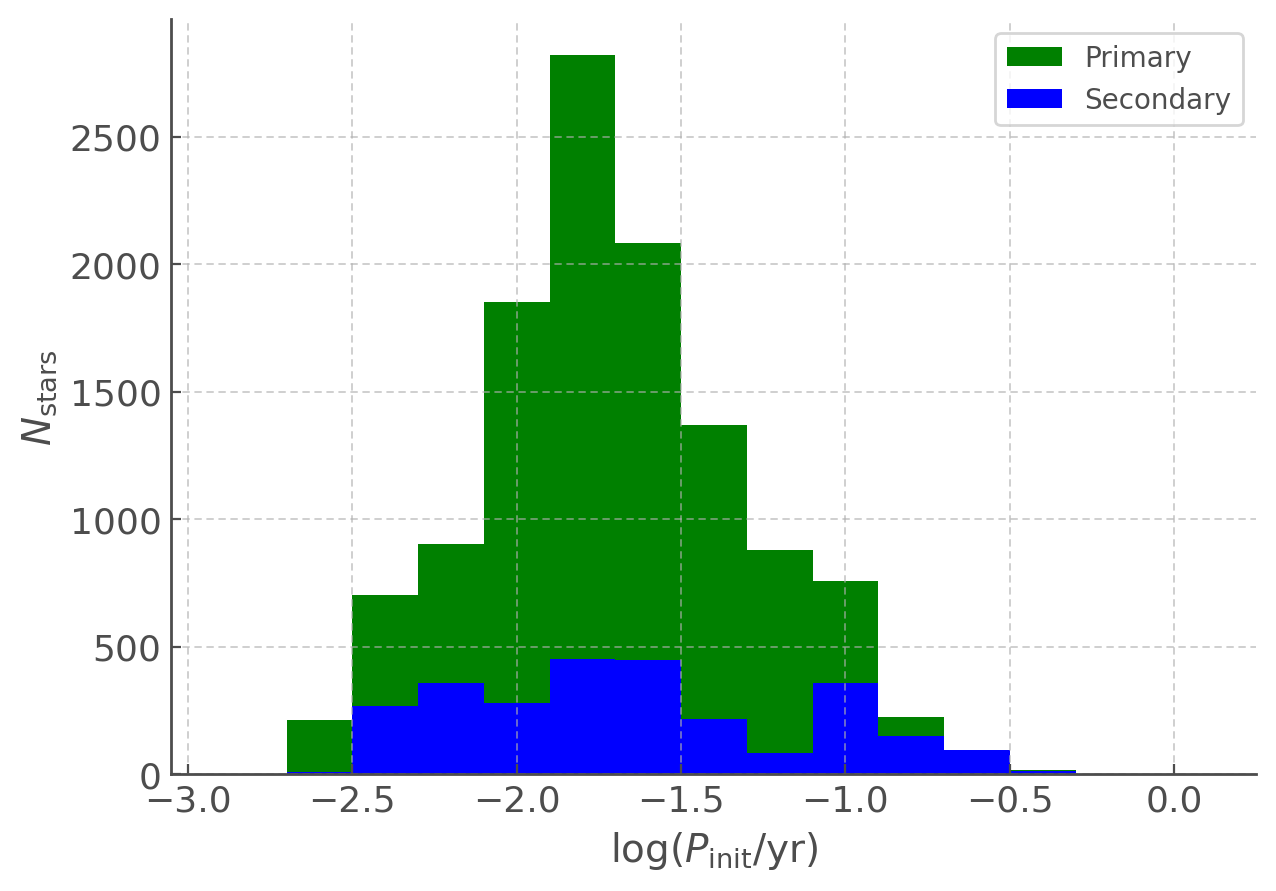}
    \caption{Distribution of initial orbital periods for BLAPs expected in the Milky Way, given constant star formation at 3\,M$_\odot$\,yr$^{-1}$.}
    \label{fig:p_orb_dist}
\end{figure}

The distribution of initial orbital periods, shown in Fig~\ref{fig:p_orb_dist}, is sharply peaked around $10^{-1.8}\,$yr or about 6\,d, with a spread of an order of magnitude in either direction. Shorter orbital periods likely produce stellar mergers, while a longer initial period will lead to a system with insufficient mass transfer to produce a low-mass pre-WD.

\label{sec:himass}

A small number of models with initial masses between 3.2 and 60$\,\rm{M}_\odot$ pass through this region (contributing 40.4 stars in the Milky Way with $3<M_\mathrm{init}/M_\odot\le5$, 5.8 stars with $5<M_\mathrm{init}/M_\odot\le8$ and 1.0 with $8<M_\mathrm{init}/M_\odot\le60$). The most massive of these must have undergone very substantial mass loss to be identified in this parameter space. The pulsating star in such a system is likely orbiting a black hole companion and may pass through X-ray luminous accretion events during their evolution. As these are exceptionally rare pathways, we do not consider these systems in more detail.

\section{Discussion}
\label{sec:discussion}

\subsection{Evolution Pathways}
\label{sec:ev_path}

We distinguish between BLAP models that have formed through Roche Lobe Overlow (RLOF) and BLAP models that have formed through Common Envelope Evolution (CEE) by comparing the initial and final orbital separations. We find that 78.5 per cent of the BLAP systems form after stable RLOF mass transfer, while 21.5 per cent undergo CEE. This is only slightly higher than the 76 per cent of systems formed through RLOF in the models computed by \cite{Han03} for subdwarfs.

Approximately 75 per cent of our BLAP models originate as the primary component of their binary system. The companion responsible for removing the hydrogen envelope is, in most instances, a low mass Main Sequence star. The remaining 25 per cent are stars which were the initial secondary in their system, and thus their companion is a white dwarf, a neutron star, or possibly a stellar mass black hole. This ratio is different from that of hot subdwarfs where approximately half of hot subdwarf stars have Main Sequence companions \citep{ReedStiening04}, although this is an empirical result, and thus subject to observational bias. Combining the first CEE and first RLOF channels of \cite{Han03} suggests that between 85 and 94 per cent of binary hot subdwarfs should have Main Sequence companions, in the absence of observational bias. Thus the fraction of compact companions may actually be higher for BLAPs than subdwarfs and more work is needed to quantify the uncertainties on these populations.

\subsection{Properties of the companions}
\label{sec:prop_comp}
Neither the observations of \cite{Pietrukowicz17} nor \cite{Kupfer19} report detection of companions to BLAPs.
The vast majority of BLAPs should have binary companions, while in the remainder the primary has undergone a type Ia supernova, leaving the BLAP as a now-single star. We calculate that of 11\,931 stars in our canonical model, only 642 (all initially secondaries) have followed this pathway ($5$\, per cent). Non-detection of the companion thus remains an open problem to be addressed using model predictions for their properties.

\begin{figure}
    \centering
    \includegraphics[width=0.48\textwidth]{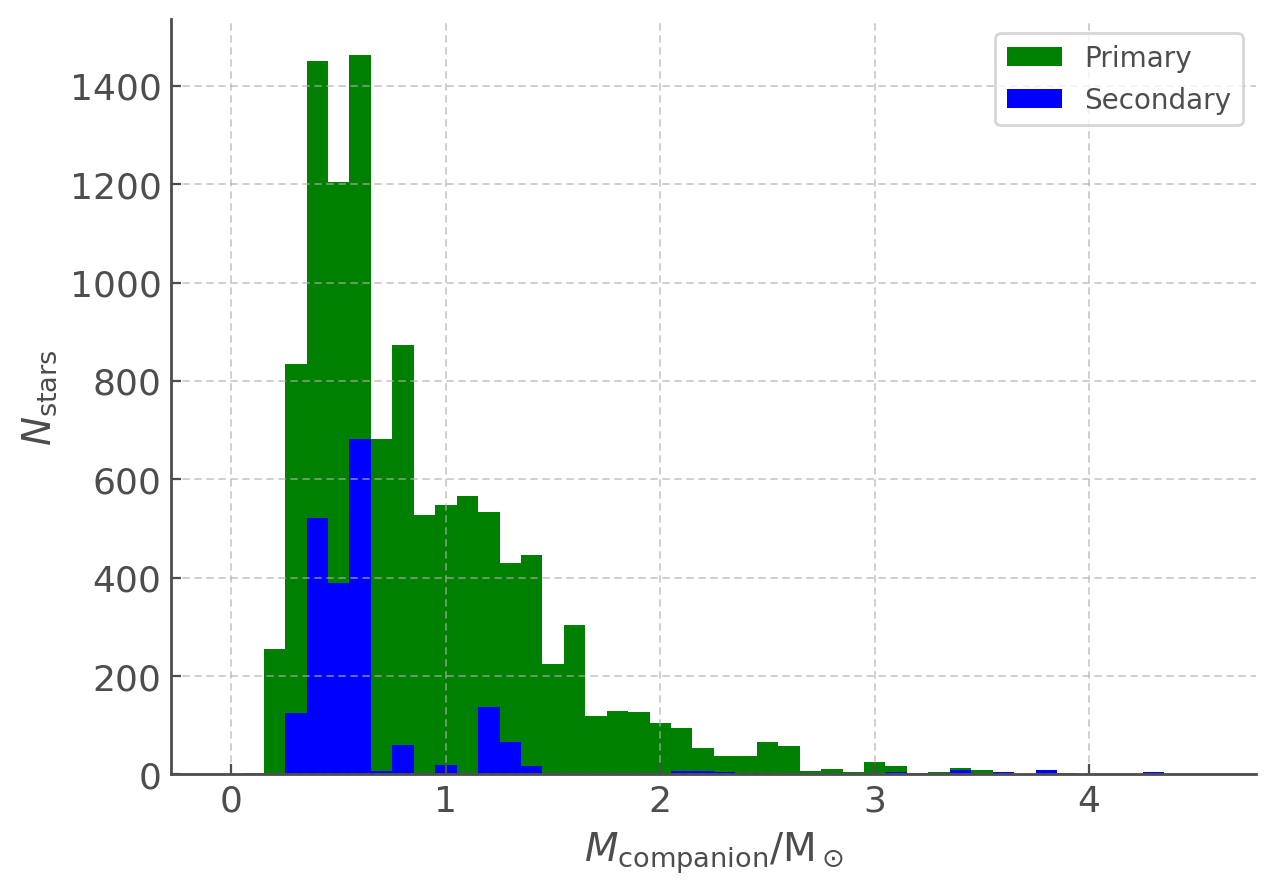}
    \includegraphics[width=0.48\textwidth]{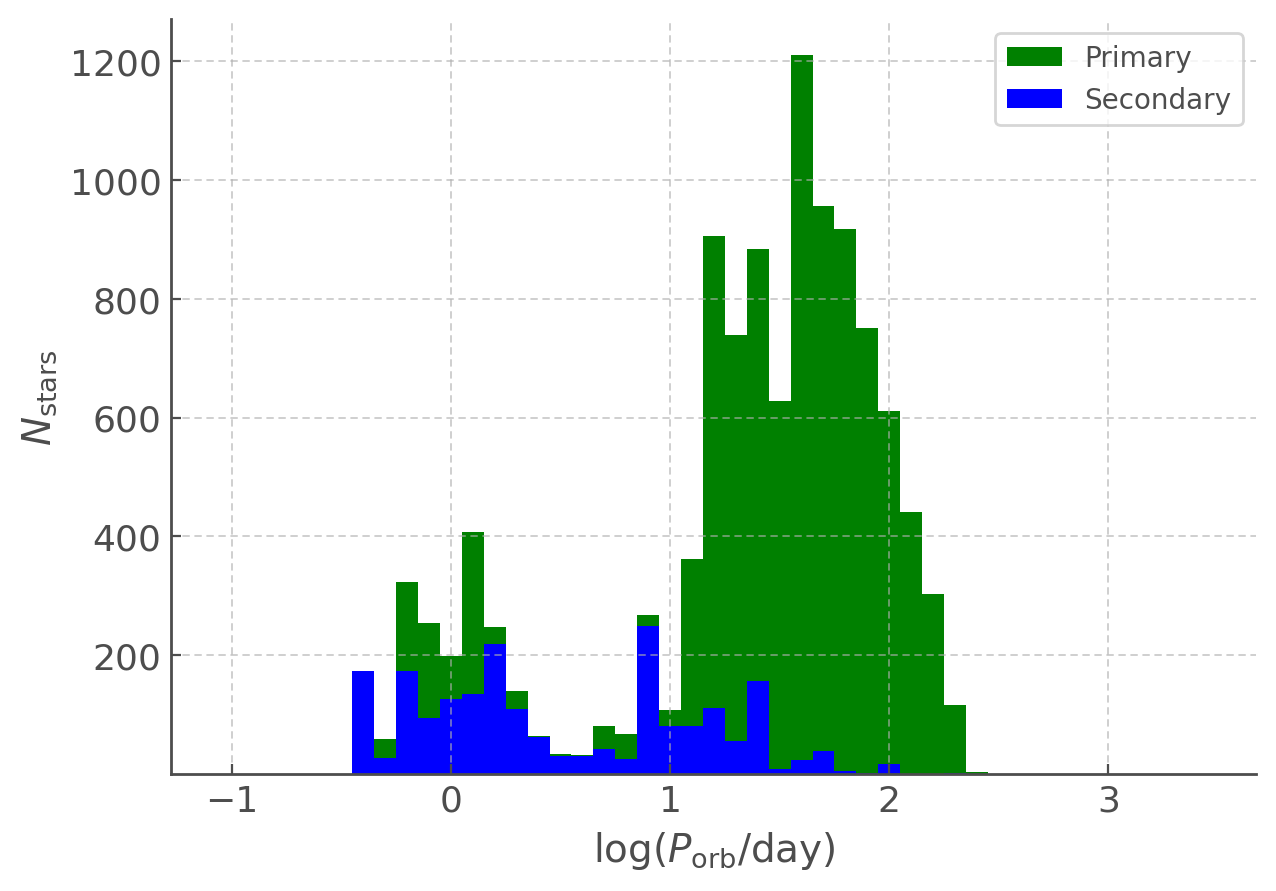}
    \caption{Distribution of BLAP companion masses (upper panel) and orbital periods (lower panel) separated according to the initial status of the BLAP progenitor.}
    \label{fig:companions}
\end{figure}

Fig~\ref{fig:companions} shows the distribution of orbital periods and companion masses of the stars when they are in the BLAP phase of evolution. The models with surviving companions are split into those where the BLAP is the initial primary and where it was initially the secondary. As Fig~\ref{fig:n_stats} demonstrates, about twenty per cent of BLAPs are expected to have evolved companions, while three quarters have low mass Main Sequence companions, and the remaining five per cent have no companions. The distribution of companion masses in the BLAP phase shows a tail extending to higher masses as a result of the secondary accreting material from the BLAP through RLOF. Nonetheless, the bulk of companions to BLAPs have a mass between $0.4$ and $0.6\,\rm{M}_\odot$. The orbital periods show a double peaked distribution. A small peak is found at about 1.2\,d and a second larger peak at an orbital period of $10^{1.6}$\,d ($\sim40$\,d). There are 1\,934 stars with orbital periods below 4.46\,d (corresponding the distribution minimum at $10^{0.65}$\,d)  and 9\,355 stars above 4.46\,d, representing about 17 per cent and 83 per cent of the systems with companions respectively.

As these orbital periods are much longer than the pulsation periods of the BLAPs, it may be possible to detect the companions through long term monitoring of these systems such as that which will be carried out by LSST or with \textit{Gaia} time-series data. Determining the nature and period distributions of companions to known BLAPs would
test the predictive powers of BPASS in the domain of low-mass, post-mass transfer systems. These companions, whether Main Sequence or evolved stars, will be faint, and hard to detect directly; Main Sequence stars that have accreted material from the BLAP during the RLOF phase are likely to be easiest to identify, as they will be the most luminous and would produce the largest radial velocity variations.

\subsection{Predicted survey number densities} 
\label{sec:survey}

\citet{2020AcA....70..241P} have recently reported the detection of 10\,000 $\delta$-Scuti variable stars from the OGLE-IV survey, drawn from the same high cadence survey of the Galactic Bulge as the much smaller sample of 14 BLAPs reported in \citet{Pietrukowicz17}. Their figure 1 illustrates the recovery limits of the sample, with very few detections at $\lvert b \rvert <2$\,degrees (where dust extinction pushes sources below the detection limit). Observations were taken in an irregularly shaped region which extended from Galactic latitudes $b=-7$ to $+5$\,degrees, and Galactic longitudes $|l|<10$\,degrees in the southern hemisphere and $-5<{l}<8$\,degrees in the north. A typical depths for recovery of variables is $I\sim19$\,mag. Observations at $b<2$\,degrees were of more limited depth and area due to the combination of extinction and confusion. 

The BLAPs identified by \citet{Kupfer19} were also selected from a dedicated Galactic plane sub-survey within the ZTF programme. In this, individual fields were studied intensively for 1.5-3 hours on two consecutive nights. The total area surveyed this way is unclear, with the four high-gravity BLAPs reported lying at galactic coordinates $(b,l)\sim$ (229,-2), (9, -10), (25, -11) and (43, -10)\,degrees, suggesting that a tight association with the Galactic Bulge is unlikely, but consistent with a disk origin. Identification with the Galactic disk population is also consistent with the young ages identified in Figure \ref{fig:age_life}.

\begin{figure*} 
    \centering
    \includegraphics[width=0.85\textwidth]{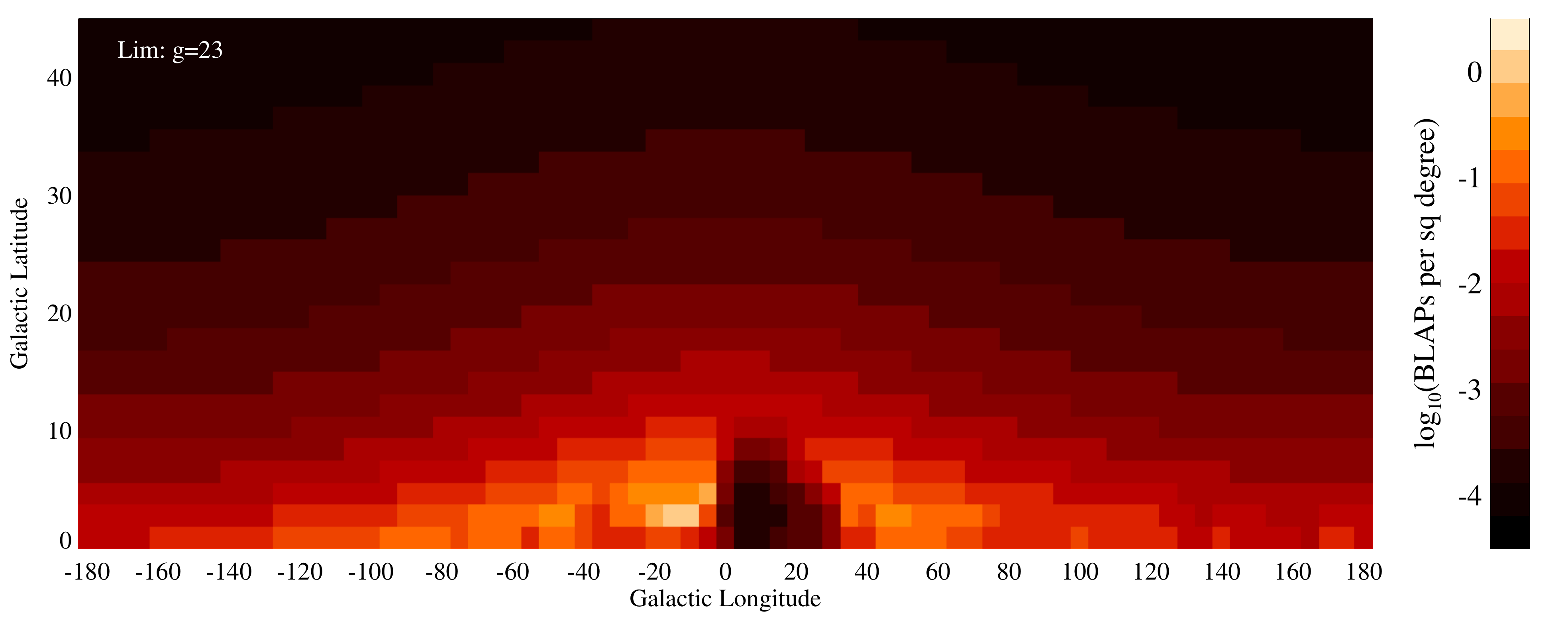}
    \includegraphics[width=0.85\textwidth]{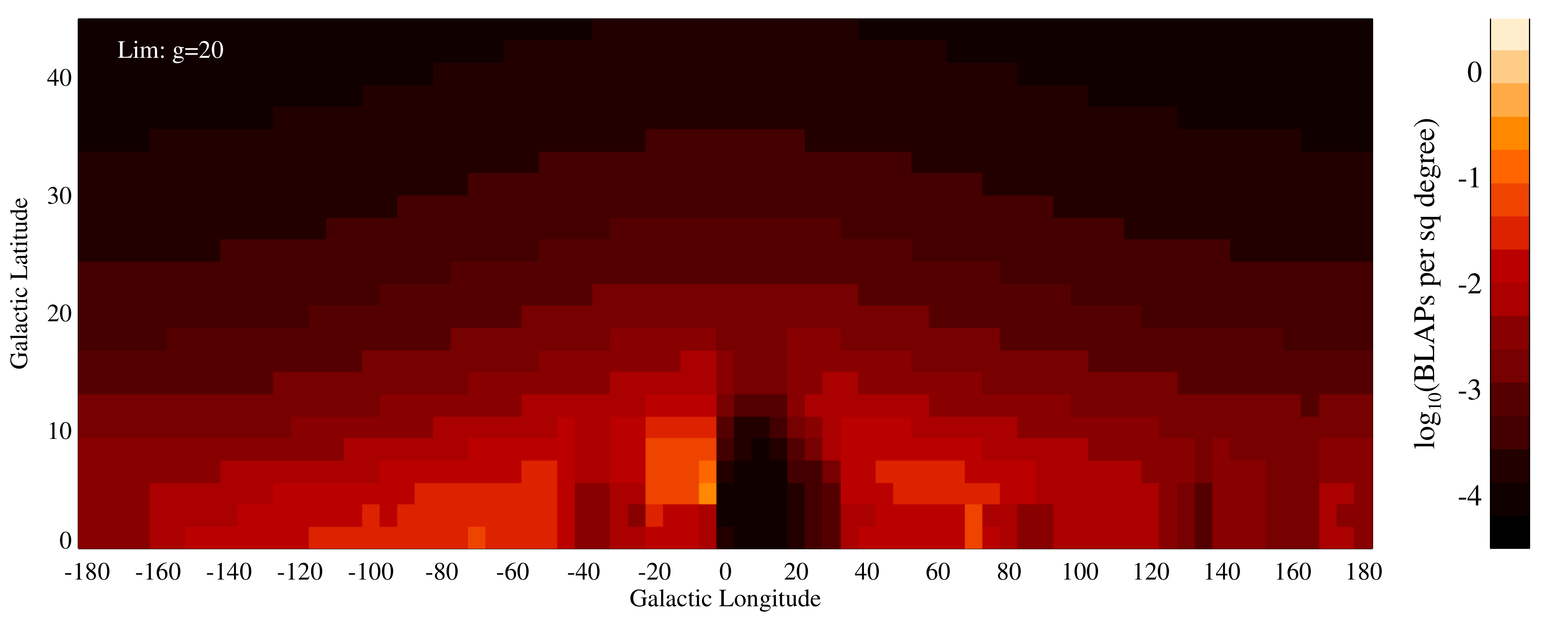}
    \caption{The expected number counts of BLAPs per square degree in all-sky surveys to depths of $g=20$ and $g=23$. Sources are modelled as having $M_g=4.0$ and originating in a disk with an exponential scale height of 300\,pc and exponential scale length of 3\,kpc. Galactic extinction is calculated as a function of sightline and distance using the model of Amores \& Lepine (2005).}
    \label{fig:galaxy_models}
\end{figure*}

In Fig \ref{fig:galaxy_models} we use our estimate of the total Galactic BLAP population to consider the number density of these sources expected per square degree in wide-area all-sky time domain surveys. BLAPs are assumed to be distributed in a disk population with an exponential scale height of 300\,pc and exponential scale length of 3\,kpc (i.e. a thin disk population) with a total mass of $3.8\times10^{10}$\,M$_\odot$, built through continuous star formation over 12.6\,Gyr and hence a total BLAP population matching our canonical model. To calculate the observable number counts, we combine this stellar density model with the dust extinction tool of \citet{2005AJ....130..659A}\footnote{http://www.galextin.org}, which estimates the Galactic extinction as a function of distance along any line of sight specified by Galactic longitude and latitude. We use the spiral galaxy (S) model and show only the northern galactic hemisphere, although the southern hemisphere is very similar.

{We assume BLAPs to have a $g$-band absolute magnitude $M_g=4.0$ \citep[typical of the sample characterised in ][]{2018A&A...620L...9R} and we consider two plausible survey depths. A limit of $g=20$ is the typical $5\,\sigma$ depth of individual observations in relatively high cadence ($\sim$ few minute to few day interval) all-sky surveys designed for transient follow-up, such as the Gravitational Wave Optical Transient Observer \citep[GOTO,][]{2020SPIE11445E..7GD}. A limit of $g=23$ approximates the depth reached at five per cent photometric uncertainty on individual ($2\times15$\,s) visits from the Vera Rubin Observatory Legacy Survey of Space and Time (VRO LSST). This survey will eventually build up many such visits (approximately 180 in 8 years, although it is not yet clear whether this general strategy will be adopted for the Galactic Plane), although these will be sparsely sampled in the time domain and a higher photometric precision is thus necessary to extract BLAP candidates within the first few years of observations.}

{Unsurprisingly for a disk population, the number density of BLAPs is expected to be heavily concentrated towards the Galactic Plane, with the highest predicted source densities in long disc sightlines passing close to the galactic centre. However extinction also plays a major role in determining number counts at a given magnitude limit, making it challenging to recover these intrinsically faint objects at very low galactic longitude without deep wide area surveys, which may be challenging in densely crowded regions. At $g=20$, the richest lines of sight through the Milky Way disk reach a maximum of 0.3 BLAPs deg$^{-1}$, while the median sightline at a galactic latitude of zero has just 0.001 BLAP deg$^{-1}$. Above galactic latitudes of $b=15$\,degrees, the number density of BLAPs is very low, such that hundreds or thousands of square degrees would have to be surveyed to this depth to identify a single target.}

An additional challenge that selections from sparsely sampled time series data may face is one of distinguishing BLAPs from the much more common hot subdwarf population. Despite limiting the upper mass limit of our models (criterion iv in section \ref{sec:bpass}), the number of CO core models discarded as unlikely to satisfy BLAP pulsation criteria outnumbered the He-core sample discussed in section \ref{sec:cands} and \ref{sec:stats} by a factor of around forty. Allowing for a broader mass range, we would expect still more higher mass systems to pass through this region of temperature-gravity parameter space. 

As a test of the predictive power of our toy model, we assess whether it reproduces the known sample. Approximating the rather complex OGLE survey area of $\sim$172\,degree$^2$ as a simple rectangle with $|l|<10$\,degrees and $2<|b|<7$\,degrees, we 
estimate the number of BLAPs recoverable in this survey to a depth of $g\sim19$ as $N=11\pm$3, consistent with the observed number ($N=14$) given uncertainty in the precise footprint, depth and other observational selection effects.

\section{Uncertainties on the Model Population}
\label{sec:uncertainties}

\subsection{Metallicity and Pulsation properties}
\label{sec:z}

The models presented in this paper have been calculated at Solar metallicity. However the observed population will probe the full range of metallicities present in the Milky Way stellar population. While the range of metallicities in the Galactic disk is more limited than that in the Bulge, it is nonetheless appropriate to consider the scale of uncertainties arising from this simplification.

The region of interest defined in Section~\ref{sec:selecting} was based on the pulsation analysis of \cite{Byrne20}. This work primarily looked at a single value of hydrogen envelope mass ($3\times10^{-3}\,\rm{M}_\odot$) and a single metallicity. Varying these parameters will alter the lifetimes of some models. Stellar metallicity, and how it evolves over time, is likely to play a more significant role than envelope thickness, as stars at lower metallicities might not be pulsationally unstable in this region at all. Searches by the OGLE team in the Magellanic clouds have found no BLAPs \citep{Pietrukowicz18} which could suggest an absence of pulsation driving, due to the reduced metal content, although the number statistics remain small and observational uncertainties may be significant.

\begin{figure}
    \centering
    \includegraphics[width=0.48\textwidth]{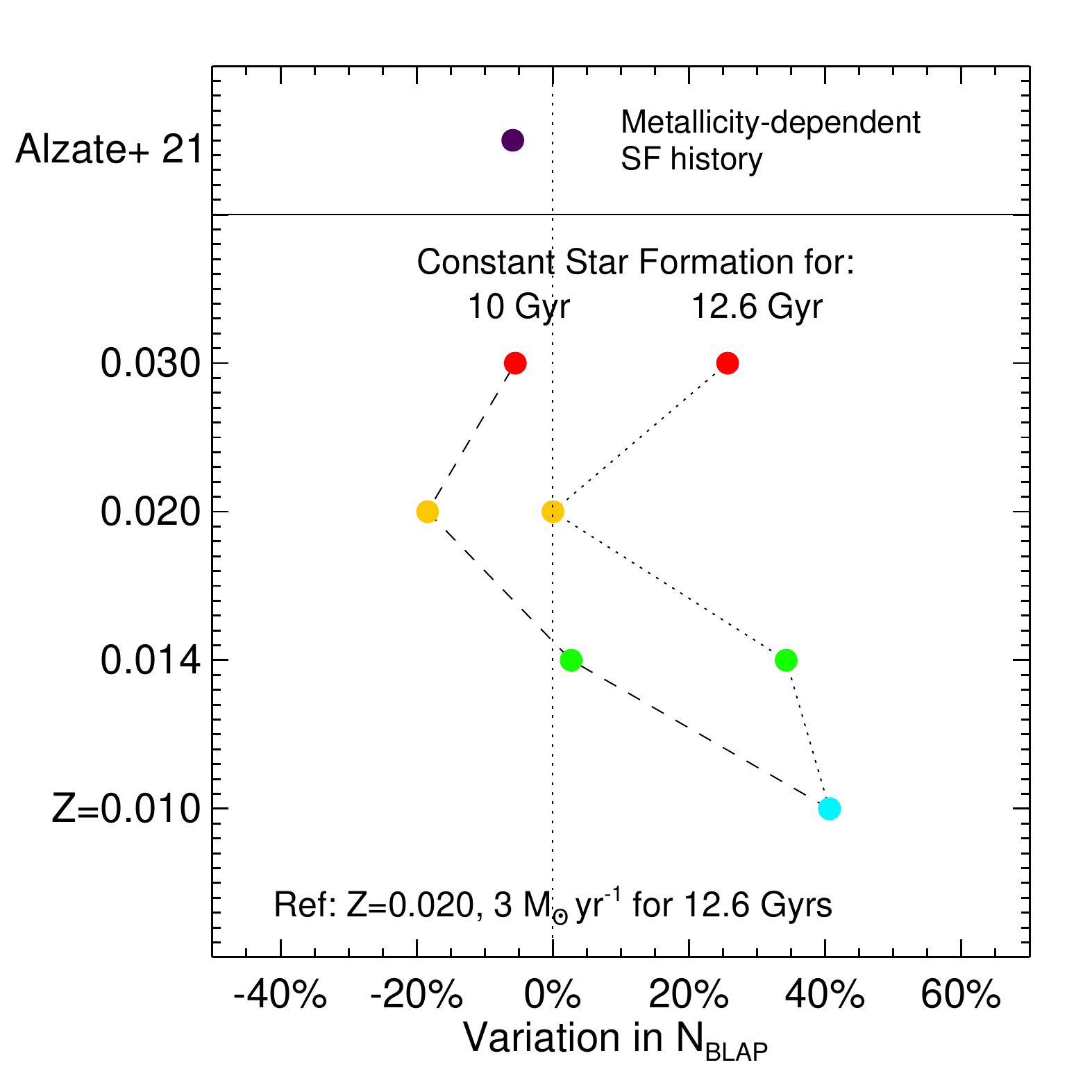}
    \caption{Variation of predicted cumulative number counts of BLAPs with age and  metallicity. Values have been scaled relative to our reference star formation history and show how the ratio in total number of BLAPs changes with length of the ongoing star formation episode and with the presumed metal enrichment of the disk population. We also show a prediction for the \citet{2021MNRAS.501..302A} stepwise stellar population model for the Milky Way as discussed in section \ref{sec:sfh}. The dotted line indicates constant star formation for log(age/years)=10.1, while the dashed line assumes the same star formation rate but for just 10\,Gyr.}
    \label{fig:metallicity}
\end{figure}

While bearing this caveat in mind, we repeat our analysis at three alternate metallicities for which stellar models are available in the BPASS model suite: $Z=$0.010, 0.014 and 0.030 (0.5, 0.7 and 1.5 Z$_\odot$ respectively). Stars with these compositions are all present in the disk population \citep[e.g.][]{2021MNRAS.501..302A}. The primary star weighting in the population synthesis at each metallicity is identical to that at $Z=0.020$ (we assume no dependence of IMF or binary initial parameters on metal content) but the age at which stars enter the BLAP region of interest, their BLAP lifetimes, their masses and the weightings of secondary models are all calculated from the results of detailed stellar evolution code and so may vary due to metal-dependent opacities, winds and other evolutionary processes.

Figure \ref{fig:metallicity} compares the total number of BLAPs expected given the same star formation history (constant star formation for 12.6\,Gyr, dotted line) at these metallicities to that at Solar metallicity, assuming the pulsational instability region is unchanged. Varying the metallicity away from $Z=0.020$  increases the expected number counts by twenty to forty per cent, regardless of whether metallicity is raised or falls.  This suggests that our number count estimates may in fact be lower limits, although detailed pulsation modelling of these systems is required in future work to accurately determine whether all these stars would pulsate. At super-Solar metallicity stellar winds are stronger and stars more efficiently stripped on the giant branch, while at lower metallicities stellar lifetimes are shorter and the probability of binary interaction higher.

\subsection{Star Formation History}
\label{sec:sfh}

In our statistics to this point we have considered a simplified model for the Milky Way disk in which star formation has persisted at a constant rate for 12.6\,Gyr (log(age/years)=10.1, $z_\mathrm{form}=4.7$) at a fixed metallicity. Such a toy model is illustrative and allows us to consider the overall properties of the population but clearly represents a substantial simplification of the complex star formation and chemical evolution histories of the Milky Way. Again, comparison to observed number counts requires the uncertainties introduced by this model to be evaluated.

Fig.~\ref{fig:metallicity} also illustrates the potential impact of the simplified Milky Way model on the predicted number counts. The dashed line indicates the percentage variation in number counts if stars in the Milky Way disk have instead been forming for only 10\,Gyr (consistent with in-situ formation following a last major merger at $z\sim1.7$ rather than constant formation since $z=4.7$). At near-Solar metallicities, the predicted number of BLAPs drops by between 20 and 25 per cent relative to counts for the same metallicity at 12.6\,Gyr. This variation is dominated by the difference in total stellar mass produced over the star formation epoch, although there is some metallicity dependence. By contrast, at half-Solar metallicity, the predicted number count remains unchanged, despite the change in total stellar mass, indicating that very few low metallicity systems enter or persist in the BLAP region of interest at ages above 10\,Gyr. 

In the same figure, we also consider a more complex case, in which the Milky Way disk population is modelled as a heterogeneous mix of stars of different metallicity, each metallicity population exhibiting a different star formation history. To do so we adopt the step-wise determination of Milky Way disk star formation history determined from \textit{Gaia} data by \citet{2021MNRAS.501..302A}. For each age bin in the star formation history model we identify the closest age bin on the BPASS model grid, and we use their mass fraction estimates as a function of age and metallicity at $Z=0.010$, 0.014, 0.017 (applied to BPASS at $Z=0.020$) and 0.030. The total stellar mass associated with the disk is then scaled to match that formed in our reference model. The resultant prediction for number of BLAPs is shown as a purple point in Fig.~\ref{fig:metallicity}. It differs from our reference model prescription less than 5 per cent.

\section{Conclusions}\label{sec:conc}

Using the BPASS detailed population synthesis code, a number of binary star configurations were identified in which a star will form a low-mass helium-core pre-white dwarf, consistent with the evolutionary picture of a BLAP. Our principle conclusions can be summarised as follows:
\begin{enumerate}
    \item  They can form from stars with a wide range of ZAMS masses, although the population synthesis weighting strongly prefers those between 1 and 2$\,M_\odot$.  Initial orbital periods range from around 1\,d to 400\,d, sharply peaked at around 6\,d, with the full range of initial mass ratios present. 
    
    \item The distribution of BLAP masses suggests that there is a double-peaked distribution, providing an indication that the observed gap between the high-gravity and low-gravity populations of BLAPs can be at least partly explained by the smaller number of BLAPs expected in the mass range between those estimated for the two groups. 
 
   \item In terms of locating companions, most companions are low mass MS stars or WDs, and are thus likely to be quite faint. The distribution of the present orbital periods has two peaks, a small one at around 1.2\,d and a larger one at around 40\,d. These are considerably longer than the pulsation periods, and may be detectable in radial velocity variations. 
 
    \item Analysis of the time each model spends in the BLAP instability regime, and the corresponding population synthesis weighting were used to calculate the number of BLAPs expected to exist in the Milky Way at the present time. The total amounts to approximately 12\,000 BLAPs.

\end{enumerate}

We use our predictions of the BLAP population to predict the density of BLAPs in the context of large-scale photometric surveys such as GOTO and VRO LSST, considering both a simple constant star formation history model and  a more complex, metallicity dependent history. Varying metallicity may affect the number of BLAPs predicted by up to around 40 per cent. Unsurprisingly, we find that the highest number density of these sources is expected to be along the galactic plane, although extinction, crowding and their intrinsic faintness all provide a challenge to detection.

\section*{Acknowledgements}

We thank members of the BPASS team for helpful discussions. ERS and CMB acknowledge funding from the UK Science and Technology Facilities Council (STFC) through Consolidated Grant ST/T000406/1. JJE acknowledges funding from the Royal Society Te Apar\=angi of New Zealand Marden Grant Scheme. This work made use of the University of Warwick Scientific Computing Research Technology Platform (SCRTP). This research made use of Astropy,\footnote{http://www.astropy.org} a community-developed core Python package for Astronomy \citep{astropy:2013, astropy:2018}. We thank the referee for their feedback on the manuscript.

\section*{Data Availability}

This analysis makes use of publically released data products from the BPASS project (hosted at https://bpass.auckland.ac.nz/). Data underlying plots in this paper will be available on request from the first author and will be made  available on the BPASS website.



\bibliographystyle{mnras}
\bibliography{mybib}








\bsp	
\label{lastpage}
\end{document}